\documentclass{aip-cp}

\usepackage[numbers]{natbib}
\usepackage{rotating}
\usepackage{graphicx}

 \usepackage{savesym}
\savesymbol{iint}
\savesymbol{iiint}
\savesymbol{iiiint}
\savesymbol{idotsint}
\usepackage[super]{cite}
\usepackage{graphicx}
\usepackage{subfig}
\usepackage{caption}
\usepackage{amsmath}
\usepackage{mwe}

\def\be{\begin{equation}}
\def\ee{\end{equation}}
\def\ba{\begin{eqnarray}}
\def\ea{\end{eqnarray}}

\newcommand\mba{\mathbf{a}}
\newcommand\mbb{\mathbf{b}}
\newcommand\bee{\begin{equation*}}
\newcommand\eee{\end{equation*}}
\def\be{\begin{equation}}
\def\ee{\end{equation}}
\def\ba{\begin{eqnarray}}
\def\ea{\end{eqnarray}}

\newcommand\dphi{\dot{\phi}}
\newcommand\eb{b e^{-\alpha\phi}}

\begin{document}

\title{The multi-measure cosmological model and its peculiar effective potential }

\author[aff1]{Denitsa Staicova}
\eaddress{dstaicova@inrne.bas.bg}

\affil[aff1]{Institute for Nuclear Research and Nuclear Energy \\Bulgarian Academy of Sciences, Sofia 1784, Tsarigradsko shosse 72, Bulgaria}

\maketitle

\begin{abstract}
The multi-measures model applied to cosmology has been recently shown to reproduce qualitatively the expected stages of the Universe evolution, along with some unexpected features. In this article, we continue its exploration with a detailed study of the effective potential of the model. An analysis of the limits of applicability of the effective potential show that during most of the Universe evolution, the effective potential is a very good approximation of the actual potential term. There is, however, a deviation between the two occurring in the earliest moments of the evolution, which has an important role on the behavior of the inflaton scalar field. In the studied cases, during this initial time, the inflaton is increasing in absolute value, which seems consistent with ``climbing up'' the effective potential. To investigate this behavior, we use numerical integration to find the numerical potential and we show that during the early stage of the evolution, the shape of the numerical potential is very different from that of the effective one and instead of a left plateau followed by steep slope, one observes only a slope with additional local maximum and minimum. This result demonstrates that for complicated equations of motion of the inflaton, one should not rely only on the notion of kinetic and effective potential terms to describe the problem as they may not be accurate in the entire numerical domain. 
\end{abstract}

\section{INTRODUCTION}
The two-mesures model has been developed in series of works by Guendelman, Nissimov and Pacheva \cite{ref01,ref01_1, ref01_2, ref01_3, 1407.6281, 1408.5344, 1507.08878, 1603.06231, 1609.06915} and it aims to describe the evolution of the Universe in an unified way -- from the early inflation needed to solve the well-known cosmological problems\cite{cosmo0, Linde, Debono}  to the graceful exit to the radiation and matter domination period and the following late-time inflation. Also, it is able to describe the dark matter content of the Universe trough a dust contribution to the energy-momentum tensor. The distinguishing feature of this theory of modified gravity is that its action features two or more independent volume-forms, one of which is the standard Riemmanian volume form and the other -- new metric-independent non-Riemannian volume form. While the introduction of the new non-Riemannian volume forms adds only purely-gauge degrees of freedom to the theory, and their effect is felt only through the ratio of the Riemannian and non-Riemannian measures, when coupled to scalar fields, it leads to many interesting features of the theory, such as inducing a hidden symmetry, unifying dark matter and dark energy into a dark fluid, producing a dynamically generated cosmological constant, etc. For full description and further applications, see the cited above articles.  

In our past works \cite{1610.08368, 1801.07133, 1806.08199}, we have performed numerical study on different modifications of the model. First in  \cite{1610.08368, 1801.07133} we have considered a model in which the action is coupled to one scalar field -- the darkon, and we have been able to fit numerically the parameters of the model, with the data from the Supernova Type 1a. Then, in \cite{1801.07133, 1806.08199}, we expanded this work with a model consisting of two scalar fields -- the darkon and the inflaton coupled to the action. For certain values of the parameters, this model is able to describe a Universe undergoing 3 stages -- early inflation, radiation-matter domination and slow late-time inflation. It featured also an early ultra-relativistic period due to the singularity in the equations of motion and a non-zero final value of the inflaton scalar field. We also observed that in one of the considered cases, the slow-roll parameters didn't correspond to the expected during the early-inflation phase. 

In this article, we study in detail the peculiarities of the effective potential in the multi-measures model featuring darkon and inflaton scalar fields. First, we study the inflaton equation and the limit in which it can be simplified to the standard form of the equation of motion of the inflaton. We show that this can happen only in one of the considered cases. In the other, such simplification is not possible due to terms depending on $\dot{\phi}(t)^2$ which cannot be neglected. In this case also we observe strong dependence on the darkon field. Second, we study numerically the applicability of the analytical limits in both cases. We show that the numerical potential term differs from the derivative of the effective potential only in a short period after the beginning of the integration. After this period, the effective potential is a very good approximation of the potential term of the inflaton equation. Finally, we investigate the observed numerically increase of the absolute value of the inflaton. In order to explain this phenomenon, we integrate numerically the potential term to recover the potential during the time period when the theoretical and numerical terms do not coincide. We discover that the potential in that period has significantly different shape from the theoretically expected one -- instead of a plateau, followed by steep slope, it has only a slope, with additional local maximum and minimum. This new shape suggests that instead of ``climibing up the potential'', as the effective potential definition suggests, the inflaton rolls down a slope with local maximum affecting the rate of the inflation and its duration. It also demonstrates that in the case of complicated inflaton equations of motion, one cannot use only the notion of kinetic and potential terms to explain the behavior of the inflaton, because of the terms proportional to the darkon field and its velocity, which do not fit the standard theory. 

\section{The equations of motion}
The analytical and numerical details on the model have been presented already in \cite{1806.08199}, so here we will outline only the equations needed for the discussion of the effective potential. 

In the Friedman-Lemaitre-Robertson-Walker metric the variation of the complete action $S$ with respect to darkon and inflaton fields and the first Friedman equation lead to the following system of two coupled differential equations and one cubic equation for the two scalar fields -- the inflaton $\phi$ and the darkon $u$ : 
\ba
v^3+3\mba v+2\mbb&=&0 \label{sys1}\\
\dot{a}(t)-\sqrt{\frac{\rho}{6}}a(t)&=&0 \label{sys2}\\
\frac{d}{dt}\left( a(t)^3\dphi(1+\frac{\chi_2}{2}\eb v^2) \right)+\hspace{25mm}&&\nonumber\\
a(t)^3 (\alpha\frac{\dphi^2}{4}\chi_2\eb v^2+\frac{1}{2}V_\phi v^2-\chi_2 U_\phi\frac{v^4}{4})&=&0\label{sys3}
\ea
where 
$a(t)$ is the scale factor, $U(\phi)=f_2 e^{-2\alpha \phi}$, $V(\phi)=f_1 e^{-\alpha \phi}$ and the $\phi$ subscript denotes derivative with respect to $\phi$. The parameters $M_0,\;M_1,\;M_2$ and $\chi_2$ are dynamically generated integration constants, where $\chi_2$ is dimensionless and $M_0, M_1, M_2$, $f_1$ and $f_2$ are dimensionful.

The parameters of the cubic equation are:
 $$\mba_{}=-\frac{1}{3}\frac{V(\phi)+M_1-\frac{1}{2}\chi_2 b e^{-\alpha\phi}\dot{\phi}^2}{\chi_2(U(\phi)+M_2)-2M_0}, \mbb_{}=-\frac{p_u}{2a(t)^3(\chi_2(U(\phi)+M_2)-2M_0)}.$$

The energy density is: $$\rho=\frac{1}{2}\dphi^2 (1+\frac{3}{4}\chi_2 b e^{-\alpha\phi} v^2)+\frac{v^2}{4} (V+M_1)+3 p_u v/4a(t)^3. $$

As outlined in \cite{1806.08199}, the cubic equation always has at least one {\em real} root, which can be used to integrate numerically the coupled system of differential equations (\ref{sys1}--\ref{sys3}).

\section{The effective potential}
 As shown in \cite{1609.06915}, after transitioning to ``Einstein-frame'' the model possesses effective Lagrangian of the generalized k-essence type with ``effective potential'' depending only on inflaton field $\phi$ of the form:
\begin{equation}
U_{eff}(\phi)=\frac{1}{4}\frac{(f_1 e^{-\alpha\phi}+M_1)^2}{\chi_2 (f_2 e^{-2\alpha\phi}+M_2)-2M_0}.
\label{U_eff}
\end{equation}

The asymptotic of the left plateau is $U_{-}=U_{eff}\vert_{\phi\rightarrow -\infty}=\frac{f_1^2}{4\chi_2 f_2}$ and asymptotic of the right plateau is
  $U_{+}=U_{eff}\vert_{\phi\rightarrow +\infty}=\frac{1}{4}\frac{M_1^2}{\chi_2 M_2-2M_0}$. For certain parameters, $U_{eff}(\phi)$ takes the form of two plateaus, the left higher than the right one, separated by a steep slope.

It is important to note that the effective potential has been obtained under the assumption that the scalar fields are static, i.e. they don't depend on the spacetime. As it is emphasized in \cite{1609.06915}, the transformed darkon scalar field $\tilde{u}$ is no longer static in Einstein frame. Since we rely on the effective potential to track the movement of the inflaton, we would like to know how good the approximation of the effective potential is and when it breaks down. 

We start by writing the {\bf inflaton equation} in the form:

\ba
\left( v(t)^2 A(t)+1 \right) \ddot{\phi}(t)-\frac{1}{2} v(t)^2 \alpha A(t) \dot{\phi}(t)^2+\left(3v(t)^2 A(t) H+ 2 v(t) A(t) \dot{v}(t)+3H\right) \dot{\phi}(t)\nonumber\\
-\frac{v(t)^2 f_1\alpha}{2 e^{\alpha\phi(t)}}+\frac{\chi_2 f_2\alpha v(t)^4}{2 e^{2\alpha\phi(t))}} = 0
\label{inflaton_eq}
\ea

\noindent where $A(t)=\frac{b_0 \chi_2}{2 e^{\alpha\phi(t)}}$. From here on, we will omit writing explicitly $\phi(t)$ as it is assumed everywhere that the inflaton $\phi$ and the darkon $v$ depend only on time. The dot denotes derivative with respect to time, the prime -- derivative with respect to $\phi$.

If one uses the slow-roll approximation (neglecting the terms $\sim\;\dot{\phi}^2,\;\; \dot{\phi}^3,\;\; \dot{\phi}^4$ and $A(t)$), equation (\ref{inflaton_eq}) simplifies to:

\be 
\ddot{\phi}+3H\dot{\phi}+W(\phi)=0.
\label{eq_infl}
\ee 

\noindent where $W(\phi)=-\frac{v(t)^2 f_1\alpha}{2 e^{\alpha\phi(t)}}+\frac{\chi_2 f_2\alpha v(t)^4}{2 e^{2\alpha\phi(t))}}$.   One can see that even in this simplified form, the potential term does not match the derivative with respect to $\phi$ of the effective potential, i.e. $W(\phi)\neq U_{eff}'$ . Furthermore, our numerical results show that there is no reason to ignore the above-mentioned terms, because they are not small in general, especially during early inflation.

The next step is to move out of the slow-roll approximation and instead to consider the following two limits of the inflaton equation (\ref{inflaton_eq}):  

Case 1: $v(t)^2 A(t)>>1$:
\be
\ddot{\phi}-\frac{1}{2}\alpha \dot{\phi}^2+\left(3 H+2\frac{\dot{v}}{v}\right) \dot{\phi}
-\frac{\alpha e^{-\alpha\phi}}{2A(t)}\left({f_1}-\chi_2 f_2 v^2 e^{-\alpha\phi)}\right) = 0
\label{U_iC1}
\ee

Case 2:  $v(t)^2 A(t)<<1$: 
\be
\ddot{\phi}+3H\dot{\phi}-\frac{v^2\alpha e^{-\alpha\phi}}{2}\left( f_1-\chi_2 f_2 v^2 e^{-\alpha\phi)}\right) = 0
\label{U_iC2}
\ee

In both cases, the velocity of the darkon plays a significant role. In order to make an analytical estimation, we use the asymptotic values of $v(t)$ for $p_u\to 0$ (i.e. we remove the dependence on $a(t)$ in $v$):

\be
v_p = \sqrt{\frac{f_1 e^{-\alpha \phi}+M_1-\frac{1}{2}\chi_2 b_0 e^{-\alpha \phi} \dot{\phi}^2}{\chi_2 (f_2 e^{-2\alpha\phi}+M_2)-2M_0}}
\ee

If $v_{pb}=v_p(b_0\to0)$, the connection with the effective potential becomes clear: $v_{pb}^2=\frac{4 U_{eff}}{f_1 e^{-\alpha\phi}+M_1}$.

The derivative of the effective potential with respect to $\phi$ written in terms of the effective potential and the velocity $v_{pb}$ becomes: 

\ba
&U_{eff}'=-\frac{2U_{eff}\alpha e^{-\alpha\phi}}{f_1 e^{-\alpha\phi}+M_1} \left(f_1-\frac{4U_{eff}\chi_2 f_2 e^{-\alpha\phi}}{f_1 e^{-\alpha \phi}+M_1})\right)\approx\notag\\
&\hspace{+5cm} \approx -\frac{v_{pb}^2}{2}\alpha e^{-\alpha\phi}
\left(f_1-v_{pb}^2f_2 \chi_2 e^{-\alpha\phi}\right)
\label{U_prime}
\ea 

Then, in the limit $v\to v_{pb}$, the equations reduce to: 

Case 1: 
\be
\ddot{\phi}+\frac{1}{2}\left(\frac{U_{eff}'}{U_{eff}}+\frac{\alpha e^{-2\alpha\phi}\chi_2 f_2 v_{pb}^4}{2U_{eff}}-\alpha \right)\dot{\phi}^2+3 H \dot{\phi} + \frac{1}{v_{pb}^2A(t)} U_{eff}' = 0
\label{can0}
\ee

Case 2:
\be
\ddot{\phi}+3H\dot{\phi} + U_{eff}' = 0
\label{can}
\ee

One can see that even in this limit, in Case 1, the equation retains its complicated form, most notably, its explicit dependence on $v$ and $\dot{v}$, replaced here with dependence on $U_{eff}$ and $U_{eff}'$ to emphasize that this effect is related to the  $\dot{\phi}^2$ term. Furthermore, even if we were able to ignore these terms, the potential term will still remain modified by the factor $1/v(t)^2 A(t)$. In Case 2, however, we were able to significantly simplify the inflaton equation and put it into what we call {\bf the standard form} for the equation of motion of the inflaton. 

The limit $v\to v_{pb}$ seems very restricting and not realistic, but it elucidates the connection between the stipulated effective potential (\ref{U_eff}) and the inflaton equation. To examine the relevance of the limits to the actual numerical solution, we continue with a numerical study of the equation.  

\section{Numerical study of the inflaton equation}
The system (\ref{sys1}--\ref{sys3}) has 12 free parameters: $\{\alpha,b_0, M_0, M_1, M_2, f_1,$ $f_2, p_u, \chi_2\}$, plus the choices for initial value of the variables $a(0), \phi(0)$ and $\dot{\phi}(0)$. For the reasoning behind the choice of the parameters, the reader is referred to \cite{1806.08199} where the numerical details have been explained in detail. 

In \cite{1610.08368} we have considered two cases for the parameters:

\begin{itemize}
 \item Case 1: $\chi_2 \sim 1,\;\;M_0\sim -0.04,\;\; 0<M_2<<\vert M_0\vert $ and $M_1\sim 1.5$

\item  Case 2: $\chi_2 << 1,\;\; M_0\sim -0.01,\;\; M_2=4\;\; (\gg\vert M_0 \vert) $ and $M_1 = 0.24\sqrt{2000\chi_2+10}\;\;\sim 0.76$

\end{itemize}

For those parameters, we recall that we have early inflation for $t=0.017-0.460$ in Case 1, and for $t=0.015-0.662$ in Case 2 and that in both cases, the late inflation starts at $t=0.71$.

\begin{figure}[!ht]
\includegraphics[scale=0.25]{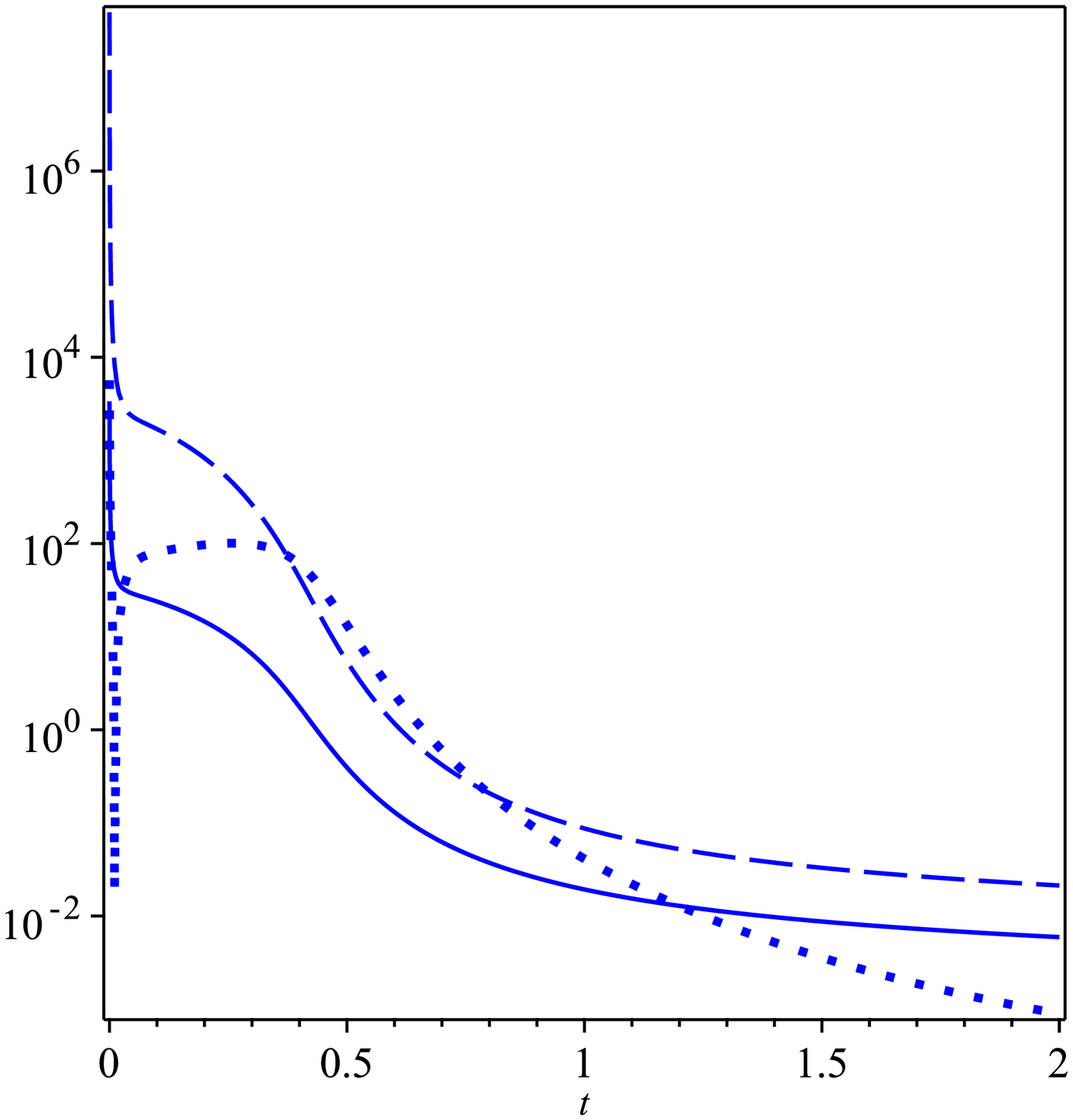}
\centering
\llap{\shortstack{%
        \includegraphics[scale=.1275]{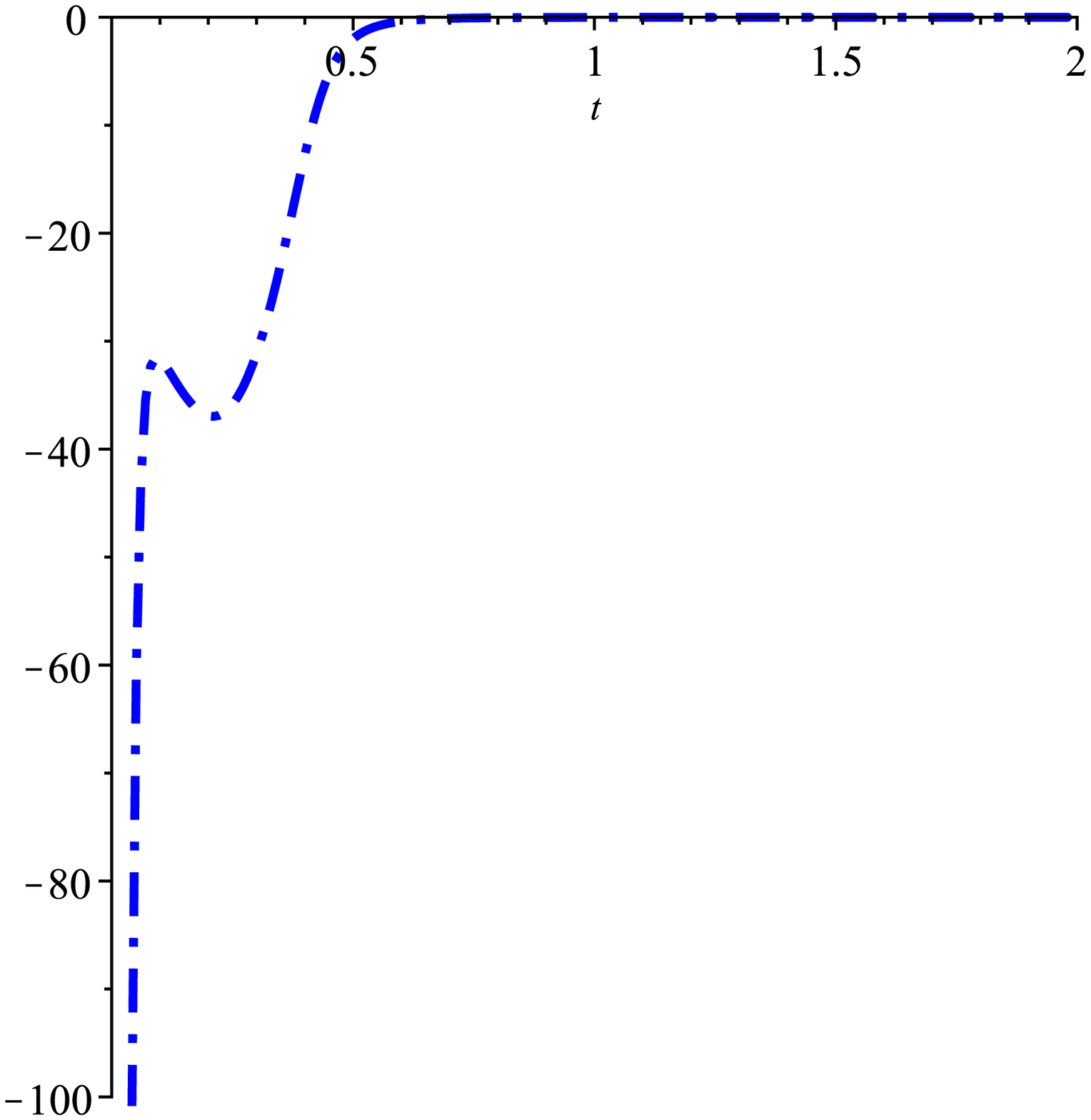}\\
        \rule{0ex}{0.8in}%
      }
  \rule{0.1in}{0ex}}
  
  \includegraphics[scale=0.25]{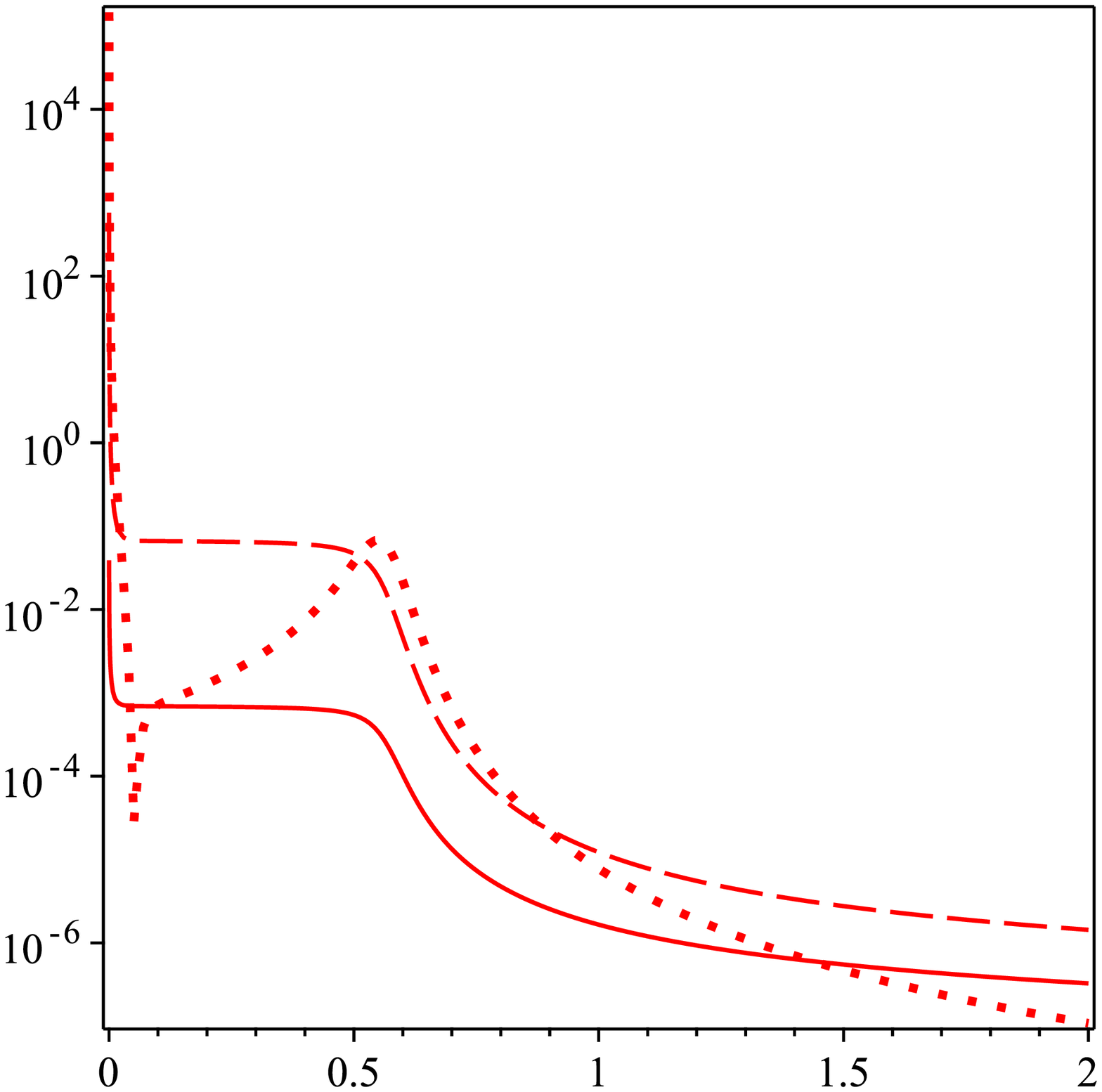}
\centering
\llap{\shortstack{%
        \includegraphics[scale=.1275]{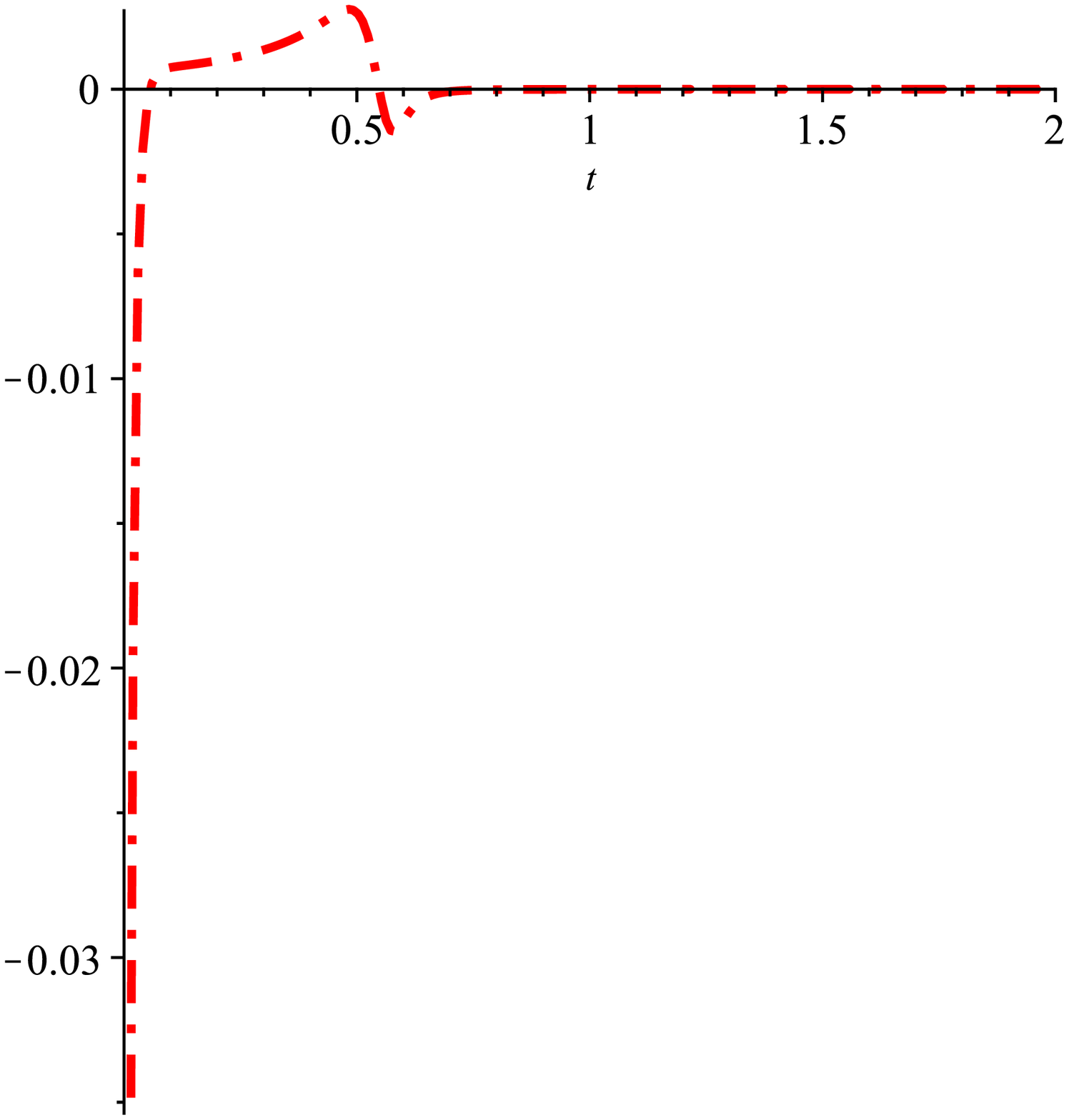}\\
        \rule{0ex}{0.8in}%
      }
  \rule{0.1in}{0ex}}
\caption{Left -- Case 1 $(\chi_2=1,\;M_0=-0.04,\;M_1=1.5,\;M_2=0.001)$
   for parameters $\{\alpha,b_0, p_u, f_1,f_2\}$ =  $\{1, 0.027, 7.7\times10^{-9}, 7, 10^{-3}\}$ .
   Right -- Case 2 in case 2 $(\chi_2=4\!\times\!10^{-5},\;M_0=-0.01,\;M_1=0.763,\;M_2=4)$. 
  for parameters $\{\alpha,b_0, p_u, f_1,f_2\}$ = 
 $\{0.64,\, 1.41\!\times\!10^{-7},\, 6.5\!\times\!10^{-24},10^{-4}, 10^{-8}\}$ . On the plots, one can see $T_1$ (solid lines), $T_2$ (dotted lines), $T_3$ (dash lines), $T_4$ (dash-dot lines).      
}  
\label{Fig1}
\end{figure}

In order to make the connection with the cases examined in the previous section, we use the numerical integration of the full system to evaluate the term $A(t)v^2(t)$ in both cases. The results are shown on Fig. \ref{Fig1}, where the solid lines correspond to this term. One can see that in Case 1, this term remains $>>1$ until almost the end of the early inflation, while in Case 2, it remains $<<1$ during the whole evolution. For this reason, one can accept that the analytical cases considered in the previous section (Equations \ref{U_iC1} and \ref{U_iC2}) correspond to the numerical cases considered here, at least until the end of the early inflation. 

Furthermore, accounting for the other parameters, in Case 2, one has $b_0\to 0, c_u\to 0, \chi_2<<1$ (to be precise $b_0\sim 10^{-7}, c_u=10^{-23}, \chi_2\sim 10^{-5}$), i.e. one can use also the limit $v\to v_{pb}$. 

In Case 1, however, we cannot assume that $v\to v_{pb}$, since $b_0 \sim 10^{-2}, \chi_2\to 1$. Therefore, it seems not possible to put Equation (\ref{U_iC1}) in the standard form Equation (\ref{can}) with a simple transformation. To eliminate the possibility that this is due to a problem of symbolic simplification, we analyze all of the terms of Equation (\ref{inflaton_eq}):

\ba
&T_1&=v(t)^2 b_0\chi_2 e^{-\alpha\phi(t)}/2 =v(t)^2 A(t) \\
&T_2&=v(t)^2 b_0 \dot{\phi}(t)^2\alpha \chi_2e^{-\alpha\phi(t)}/4 \\
&T_3&=3 v(t)^2 b_0\chi_2 H e^{-\alpha\phi(t)}\\
&T_4&=v(t) b_0\chi_2 \dot{v}(t)e^{-\alpha\phi(t)}\\
&T_5&=-v(t)^2 f_1\alpha e^{-\alpha\phi(t)}/2+\chi_2 f_2\alpha v(t)^4 e^{-2\alpha\phi(t)}/2
\ea
using the actual solutions of the full system.

From Figure \ref{Fig1} one can see that indeed, during early inflation, for Case 1 all the terms are big ($T_1, T_2, T_3, T_4>>0$), while for Case 2, all the terms are small ($T_1, T_2, T_3, T_4\to 0$). This once again confirms that for Case 2, the approximation (\ref{can}) indeed works during the whole evolution, except for the singularity in the beginning. 

\begin{figure}[!ht]
  \centering
  \begin{tabular}[b]{c}
    \includegraphics[width=.25\linewidth]{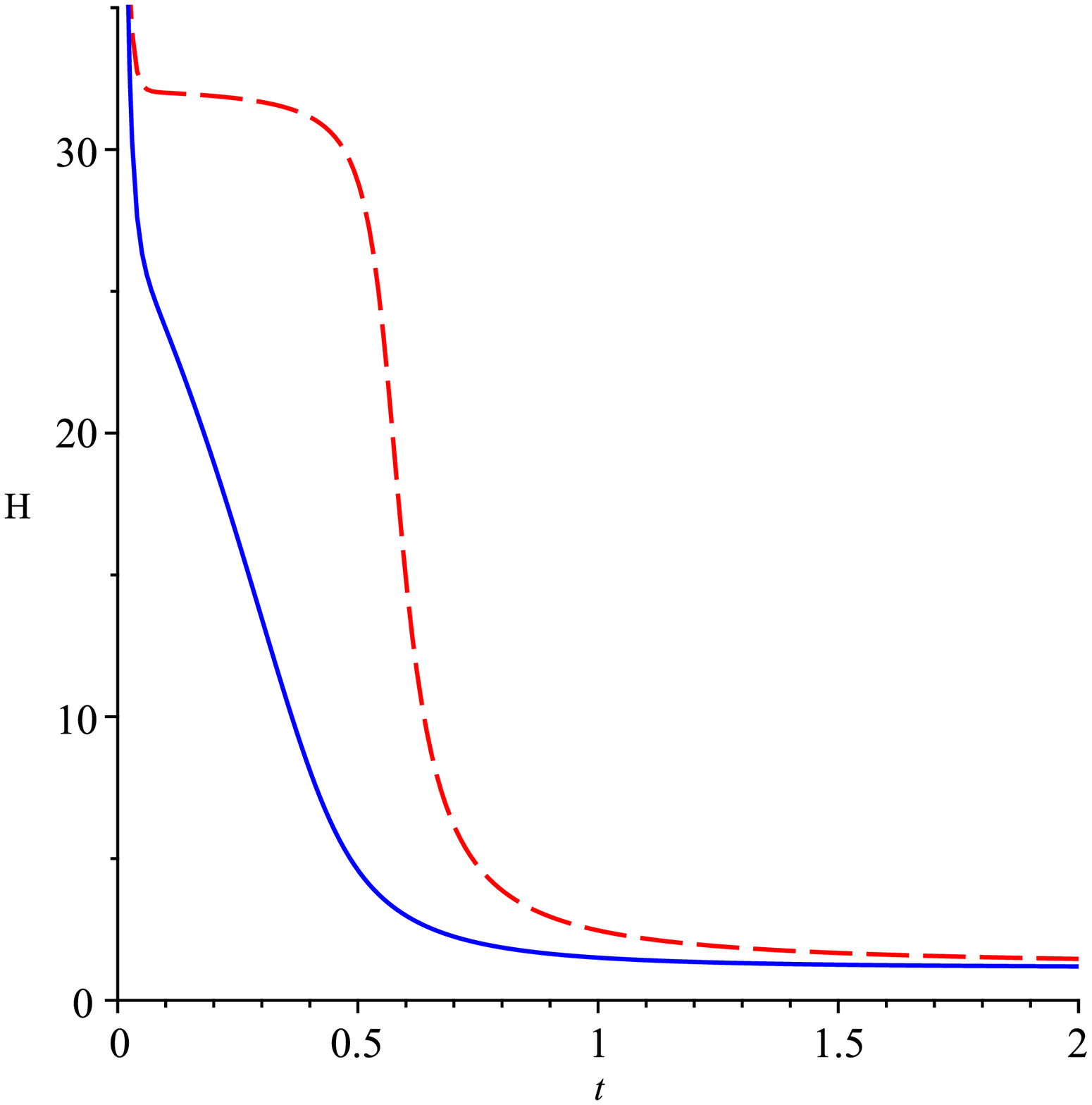} \\
    \small (a)
  \end{tabular} \qquad
  \begin{tabular}[b]{c}
    \includegraphics[width=.25\linewidth]{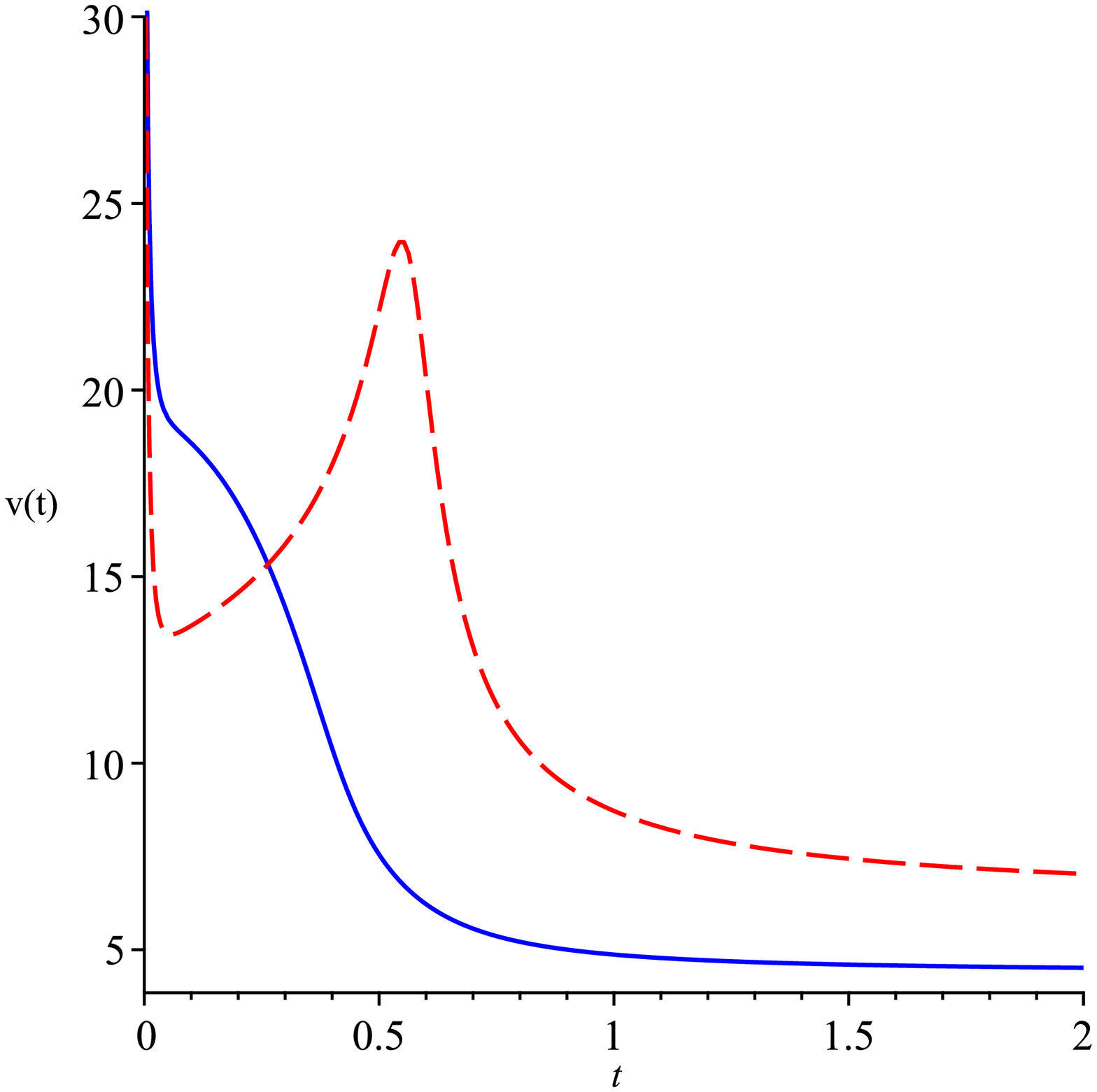} \\
    \small (b)
  \end{tabular}
  \begin{tabular}[b]{c}
    \includegraphics[width=.25\linewidth]{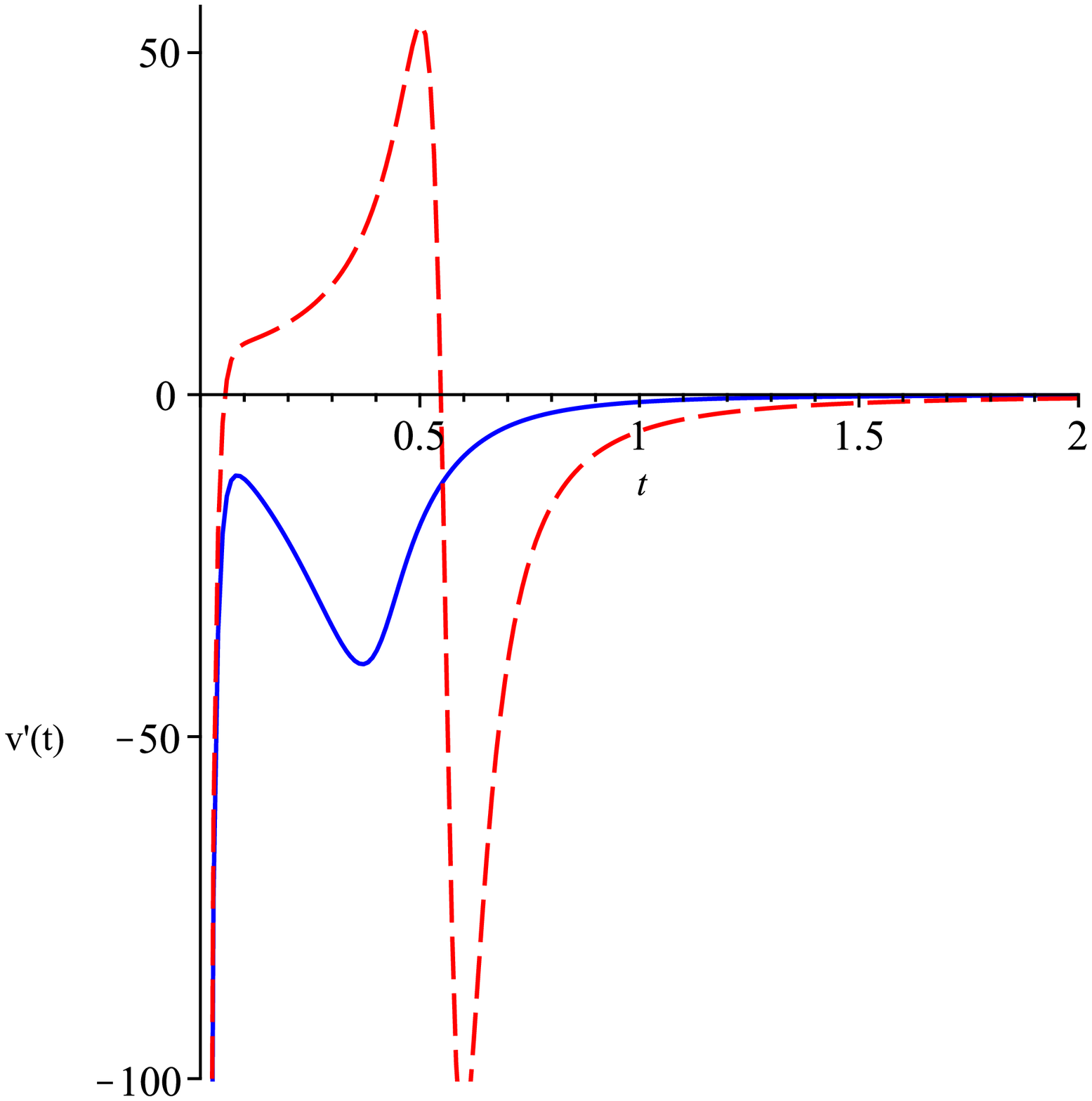} \\
    \small (c)
  \end{tabular}
  \caption{Using the same parameters in Figure \ref{Fig1}, where solid corresponds to Case 1 and dashed to Case 2, we display the Hubble parameter $H$, the darkon field $v(t)$ and its first derivative $\dot{v}(t)$ .
}  
    \label{Fig2} 
\end{figure}

On Figure \ref{Fig2} we demonstrate how the Hubble parameter $H$, the darkon $v(t)$ and the velocity of the darkon $\dot{v}(t)$ change with time. As discussed in \cite{1806.08199}, Case 2 differs from Case 1 in the length of early inflation -- it is considerably longer than for Case 1. Physically one expects extremely short initial inflation. Finding values of the parameters for which that happens, however, has so far proven very difficult and is beyond the scope of current work. 

It is clear that in both cases, one cannot neglect the darkon field or its velocity (Figure \ref{Fig2} b) and c)), especially during early inflation, when their values are big. Another interesting feature related to the darkon is that it tends to a constant during the late inflation (i.e. when $\phi\to const$), as one would expect during the current, dark-energy dominated epoch. 

\begin{figure}[!h]
  \centering
  \begin{tabular}[b]{c}
    \includegraphics[width=.3\linewidth]{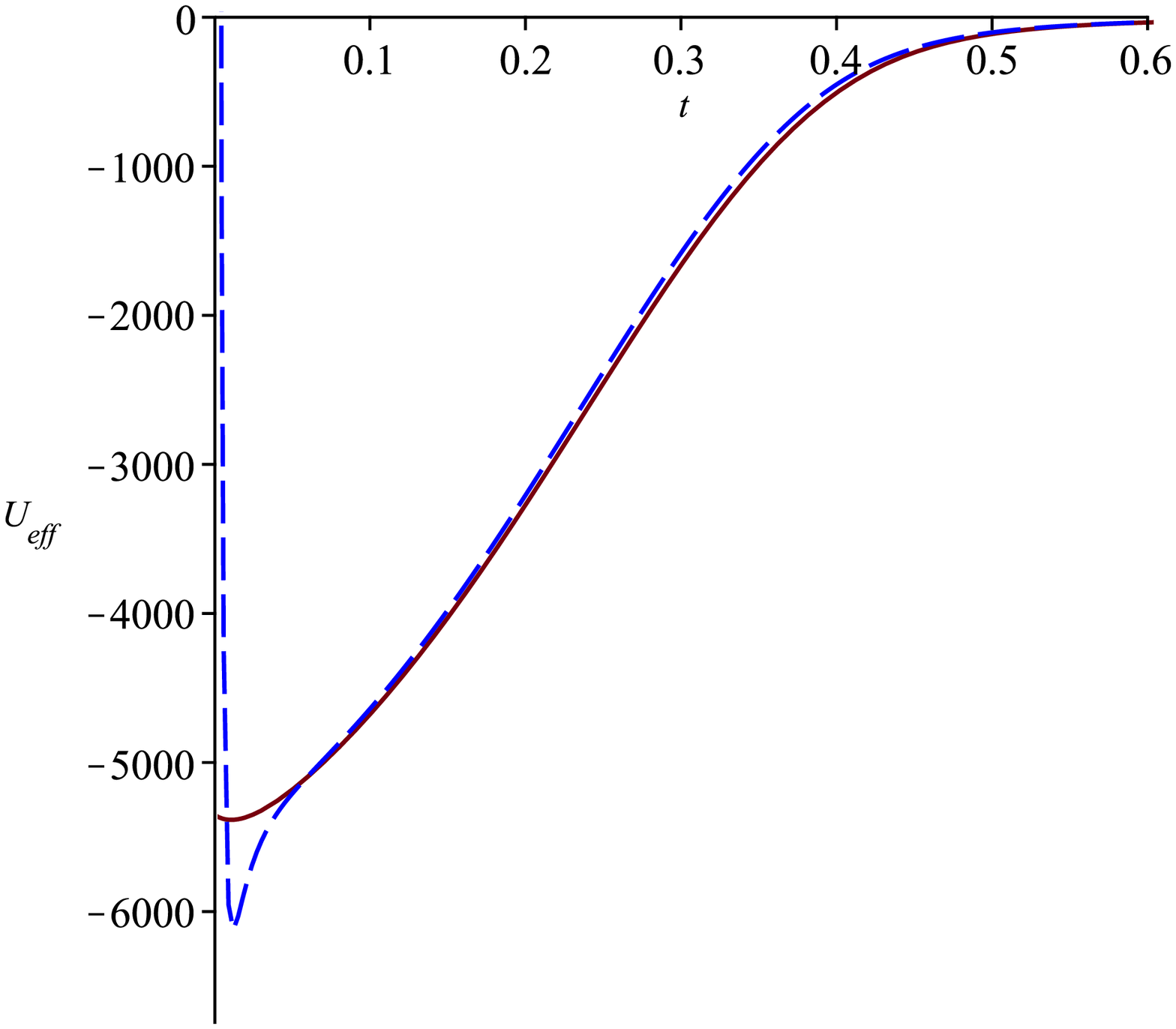} \\
    \small (a)
  \end{tabular} \qquad
  \begin{tabular}[b]{c}
    \includegraphics[width=.3\linewidth]{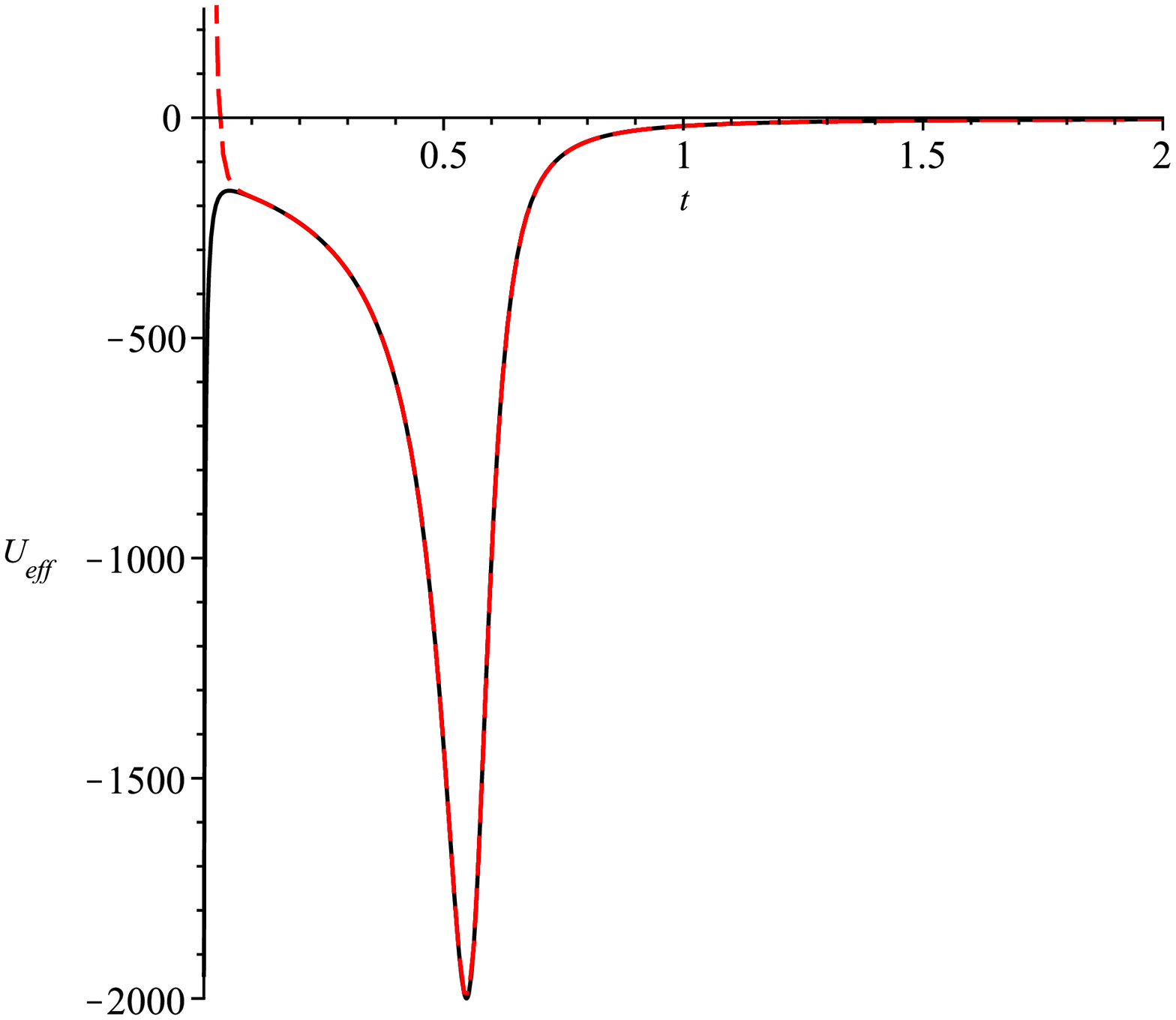} \\
    \small (b)
  \end{tabular}
 
  \caption{Using the same legend as in Figure \ref{Fig2}, we compare the term $T_5$ with the derivative of the effective potential $U_{eff}'$ in the two cases. Note, here the x-axis for Case 1 stops before $t=2$ in order to zoom on the interval in $t$ where the difference is the most significant}  
    \label{Fig3} 
\end{figure}

On Figure \ref{Fig3}, we compare the last term in Equation \ref{inflaton_eq}, i.e. $T_5=W(\phi)$, with the derivative of the effective potential $U_{eff}'$. One can see that in both cases, the two curves differ only around the initial singularity and in Case 2, they coincide almost entirely after some time. In Case 1, they differ slightly during the whole period, which confirms that the effective potential is only approximately correct. Still, this figures shows that except during the ultra-relativistic period and the early stages of the initial inflation, effectively $W(\phi)\approx U_{eff}'$. We would consider the effect of the initial deviation in the next section.

Before looking into the movement of the inflaton across the effective potential, we would like to discuss further the slow-roll approximation. We have seen in \cite{1806.08199} that both Case 1 and 2 reproduce the 3 significant stages of the Universe -- early inflation, matter-radiation domination and late inflation. Also both of them have initial ultra-relativistic regime. What we have observed, however, is that if one uses the slow-roll parameters to prove inflation, the two cases behave differently. Namely, in Case 1, the slow-roll parameters during early inflation are not much less than 1, as expected from the slow-roll regime, but in Case 2 they are. While there are inflation theories generalizing the slow-roll regime (namely the constant roll inflation \cite{1411.5021, 1702.05847, 1808.01325}), the situation here differs because the equations of motion and the effective potential follow from the Lagrangian of the theory and they can be fine-tuned only trough the choice of parameters. For this reason, the question of the applicability of the slow-roll approximation (or any other relying on an inflaton equation of the same standard form) remains open. 

One can expect that the effective potential being only approximate will affect the slow-roll parameters only in their so called potential definition. For this reason in \cite{1806.08199}, we use their kinematic definitions:
\be
\epsilon=-\frac{\dot{H}}{H^2},\;\; \eta=-\frac{\ddot{\phi}}{H\dot{\phi}}
\label{eps_eta}
\ee

However, even the kinematic definitions are not entirely free of the assumption that one works with a particular form of the inflaton equation. While the definition of $\epsilon$: 
($\frac{\ddot{a}(t)}{a(t)}=\dot{H}+H^2=H^2(1-\epsilon)$)
\noindent is purely kinematic and measures the "exponentiality" of $a(t)$, the definition of $\eta$ coming from the requirement $\ddot{\phi}(t)<<3H\dot{\phi}$ implies that one assumes the standard form for the inflaton equation (\ref{can}). In Case 1, however, one cannot obtain such standard form during early inflation, because of the additional term $2\frac{\dot{v}(t)}{v(t)}\dot{\phi}(t)-\frac{1}{2}\alpha \dot{\phi}(t)^2$ in Equation \ref{U_iC1}. For this reason, the parameter $\eta$ is not well suited for proving inflation and may show deviations when the contribution of the additional term is strong. 

Finally, one shouldn't forget that in the considered cases, one cannot assume $\dot{\phi}\to0$ during early inflation, even though we use $\dot{\phi}(0)=0$. This is because the velocity of the inflaton is not a free parameter as we discuss below. For this reason, one should not neglect terms proportional to it. 
   
\section{The motion of the inflaton}

A novel feature can be observed if one examines the evolution of the inflaton scalar field $\phi$ numerically. We see that the beginning of the early inflation coincides with the inflaton increasing its absolute value, which if one accepts the effective potential to be exact, should be interpreted as the inflaton climbing up the slope (because the integration starts on the slope). Here we focus mostly on Case 2, since the effect is much more emphasized. It is, however, also present in Case 1, pointing to a parameter-independent effect. 

First we check how the system reacts to changes in the values of $\phi(0)$ and $\dot{\phi}(0)$.

\begin{figure}[!ht]
  \centering
  \begin{tabular}[b]{c}
    \includegraphics[width=.25\linewidth]{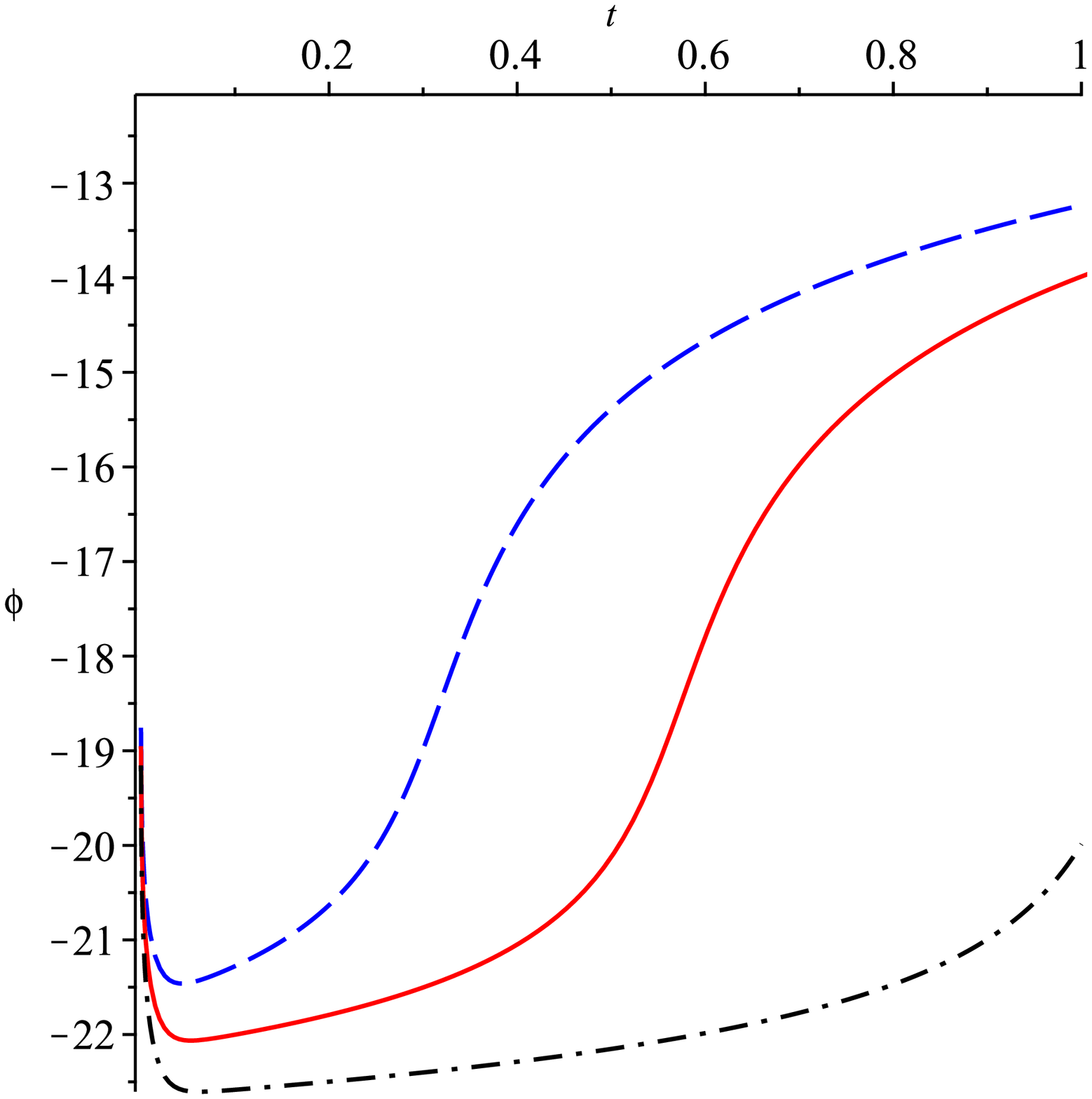} \\
    \small (a)
  \end{tabular}
  \begin{tabular}[b]{c}
    \includegraphics[width=.25\linewidth]{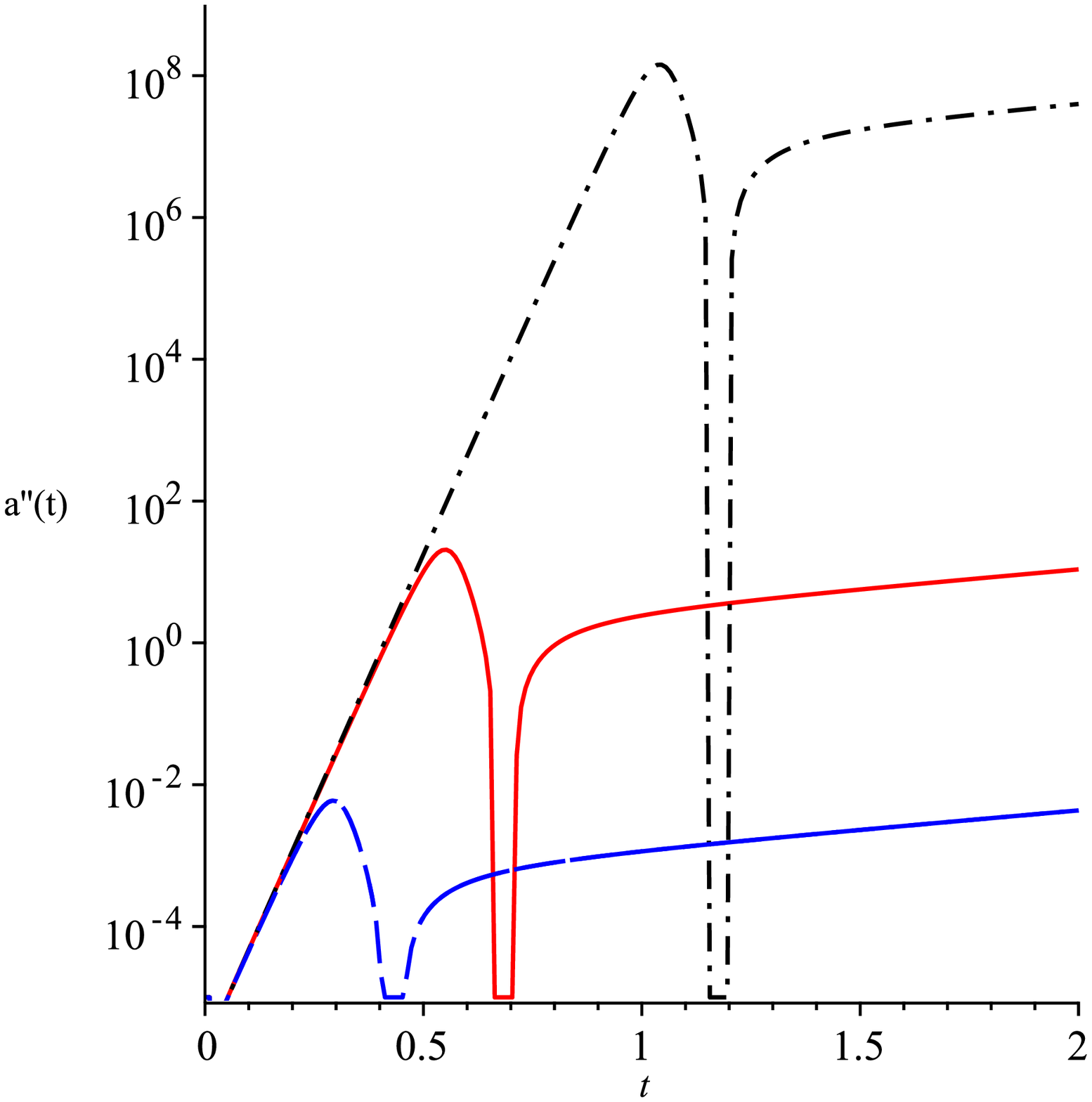} \\
    \small (b)
 \end{tabular}
 \begin{tabular}[b]{c}
    \includegraphics[width=.25\linewidth]{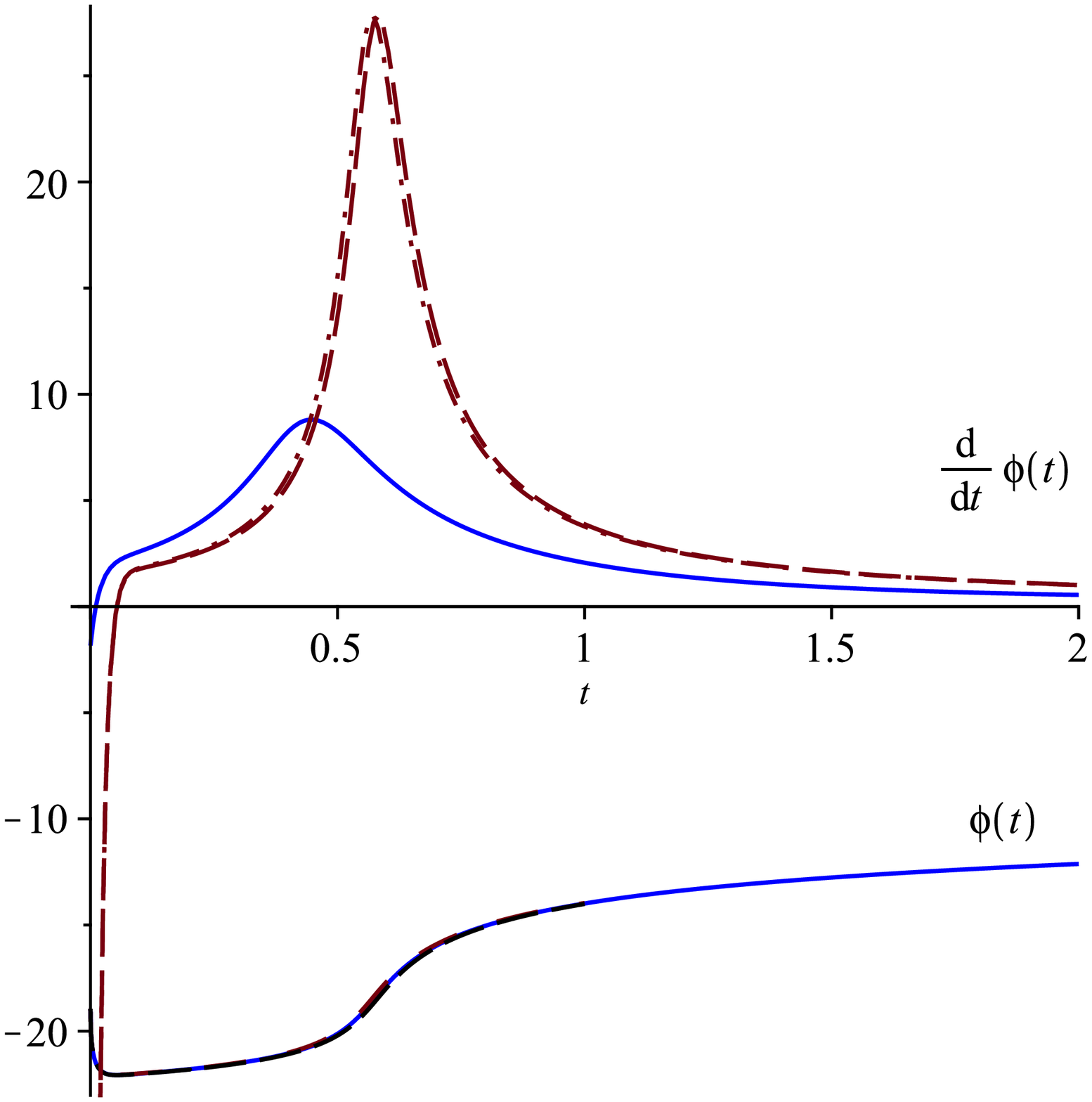} \\
    \small (c)    
  \end{tabular}
  \caption{Using the parameters for Case 2, we plot a) the inflaton $\phi$ and b) the second derivative of the scale factor $\ddot{a}(t)$ for $\phi(0)=-17.9$ (dashed line), $\phi(0)=-18$ (solid line), $\phi(0)=-18.1$ (dot-dashed line). c) the dependence$\phi(t)$ and $\dot{\phi}(t)$ for $\dot{\phi}(0)=0, \pm 10^4$ where the solid line corresponds to zero initial velocity.
}  
    \label{Fig4} 
\end{figure}

On Figure \ref{Fig4} a) we have plotted what happens with the inflaton field $\phi$ if we deviate up or down the slope from $\phi(0)=-18$, i.e. we use as initial value $\phi(0)=-17.9$ and $\phi(0)=-18.1$ (note the effective potential is the same since we are not changing other parameters). One can see the clear increase in $|\phi(t)|$ in the the 3 cases. 

Figure \ref{Fig4} b) shows the second derivative of the scale factor for the 3 cases from Figure \ref{Fig4} a). One sees that the higher $|\phi(0)|$, the longer and stronger initial inflation one observes. Note here, since it is a logarithmic plot, we have cut the negative part of the scale factor corresponding to both ultra-relativistic matter and radiation-matter domination. 

On Figure \ref{Fig4}, c) we have plotted the effect of changing the initial velocity of the inflaton, i.e.  $\dot{\phi}(0)=0, \pm 10^4$ . The effect on $\phi(t)$ is minimal, while the effect on $\dot{\phi}(t)$ is more visible. One needs not to forget, however, that the only normalized solution is the main one (the solid line). For this reason, the deviation from the default solution both in terms of $\phi(0)$ or $\dot{\phi}(0)$ lead to the corresponding changes in the zeros of $\ddot{a}(t)$, such as the one shown on Figure \ref{Fig4} b) -- it changes the timing of the 3 stages of the Universe evolution and thus it breaks the normalization ($a(1)=1$). From this figure, however, becomes clear that $\dot{\phi}(t)$ is not a free parameter but it is connected trough the equation of state (the cubic equation) with $a(t)$ and $\phi(t)$. 

Because the equation (\ref{inflaton_eq}) possesses a singularity for $t=0$, we check whether the behavior of the inflaton may be due to the initial singularity by comparing the time-scales of the early inflation and the movement ``up the slope''. We see that the inflaton increases in absolute value until $t=[0.0443, 0.0537, 0.0627]$ for $\phi(0)=[-17.9, -18, -18.1]$ respectively, while the early inflation kicks in at $t= [0.0147, 0.0145, 0.0144]$. This means that the inflaton field has still not reached its maximal absolute value when the inflation starts, i.e. this effect starts during the ultra-relativistic regime, but it becomes maximal during early inflation which is considered to be a physical regime. The time-period in which $\phi$ reaches its maximal value, however, is during the time-interval where $U_{eff}'\neq T_5$ (see Fig \ref{Fig3} b)). Thus any reference to the effective potential cannot be entirely trusted.

In summary we observe that: 
\begin{enumerate}
 \item Independent of the value of $\phi(0)$, there is a period for which $\phi$ increases in absolute value -- i.e. it climbs up the slope of the effective potential
 \item This effect becomes more pronounced the more we increase in absolute value $\phi(0)$
 \item The effect does not depend on $\dot{\phi}(0)$. That is to say that this effect is not connected with the inflaton gaining kinetic energy so that it can climb the slope.
 \item The time during which this happens puts it in the interval when the effective potential is not a good approximation of the potential term. 
\end{enumerate}

\begin{figure}[h]
 \centering
  \begin{tabular}[b]{c}
    \includegraphics[scale=0.24]{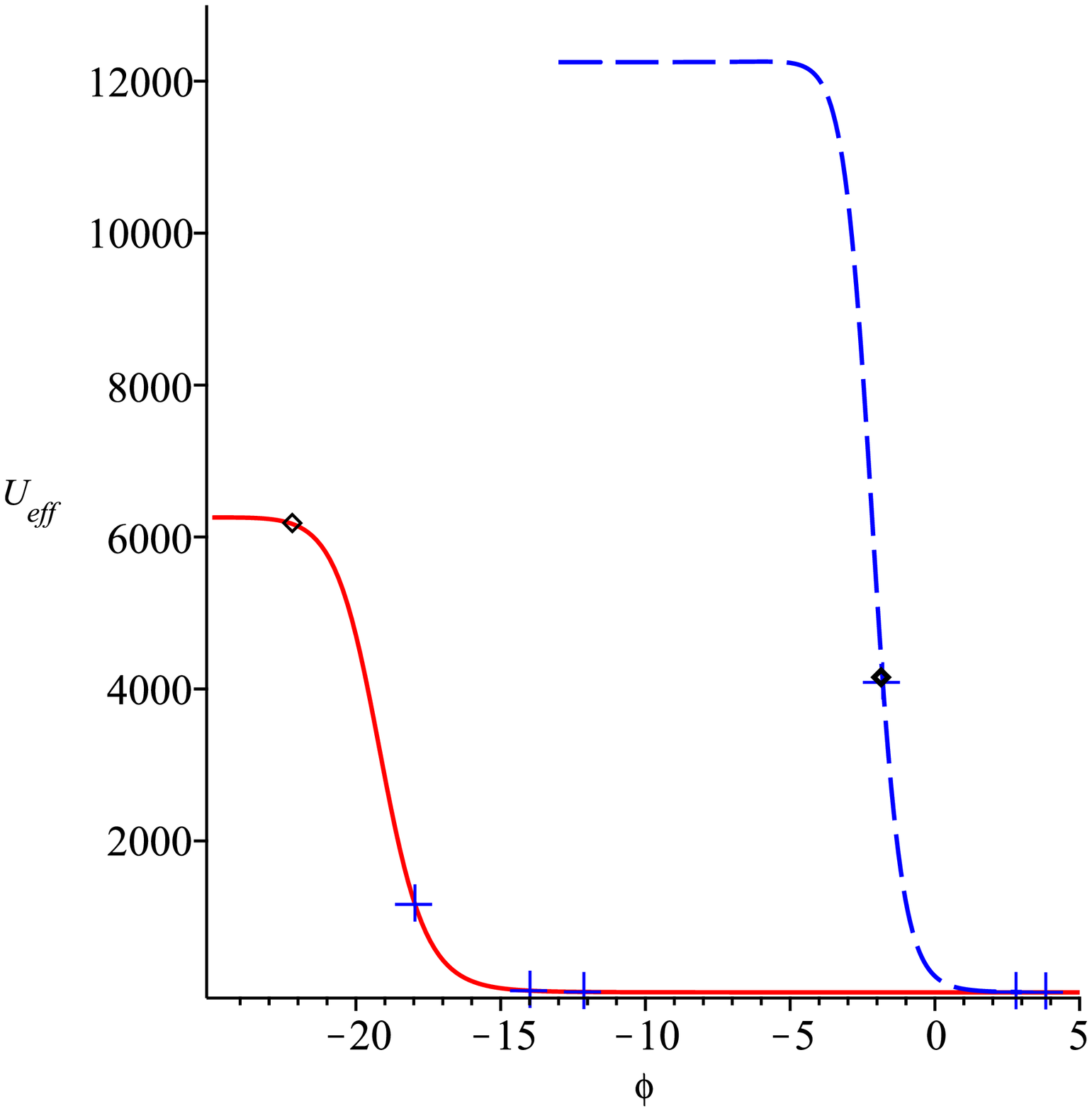} \\
      \rule{0.1in}{0ex}
    \small (a)
  \end{tabular} 
  \begin{tabular}[b]{c}
    \includegraphics[scale=0.25]{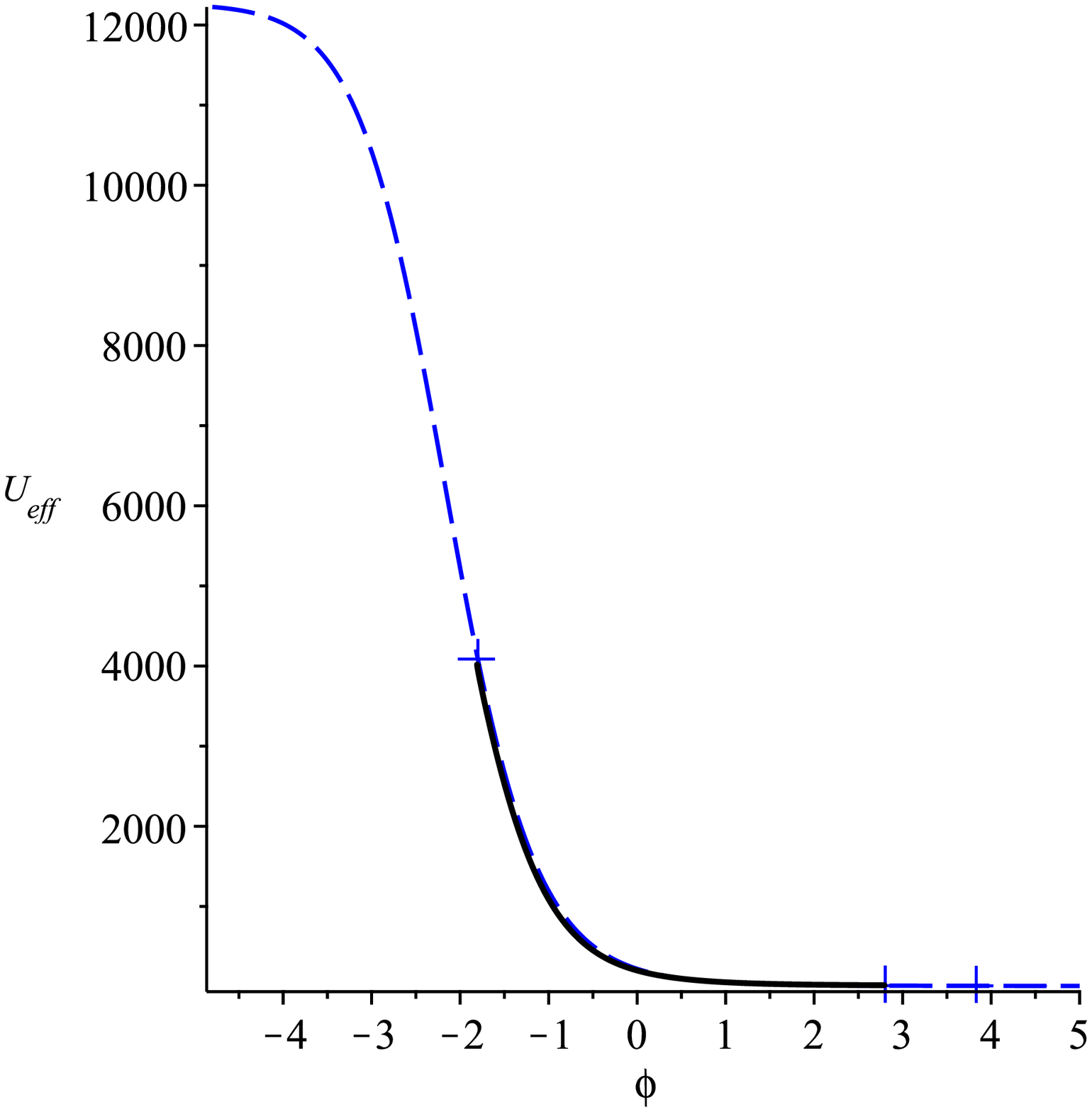}
\centering
\llap{\shortstack{%
        \includegraphics[scale=.1275]{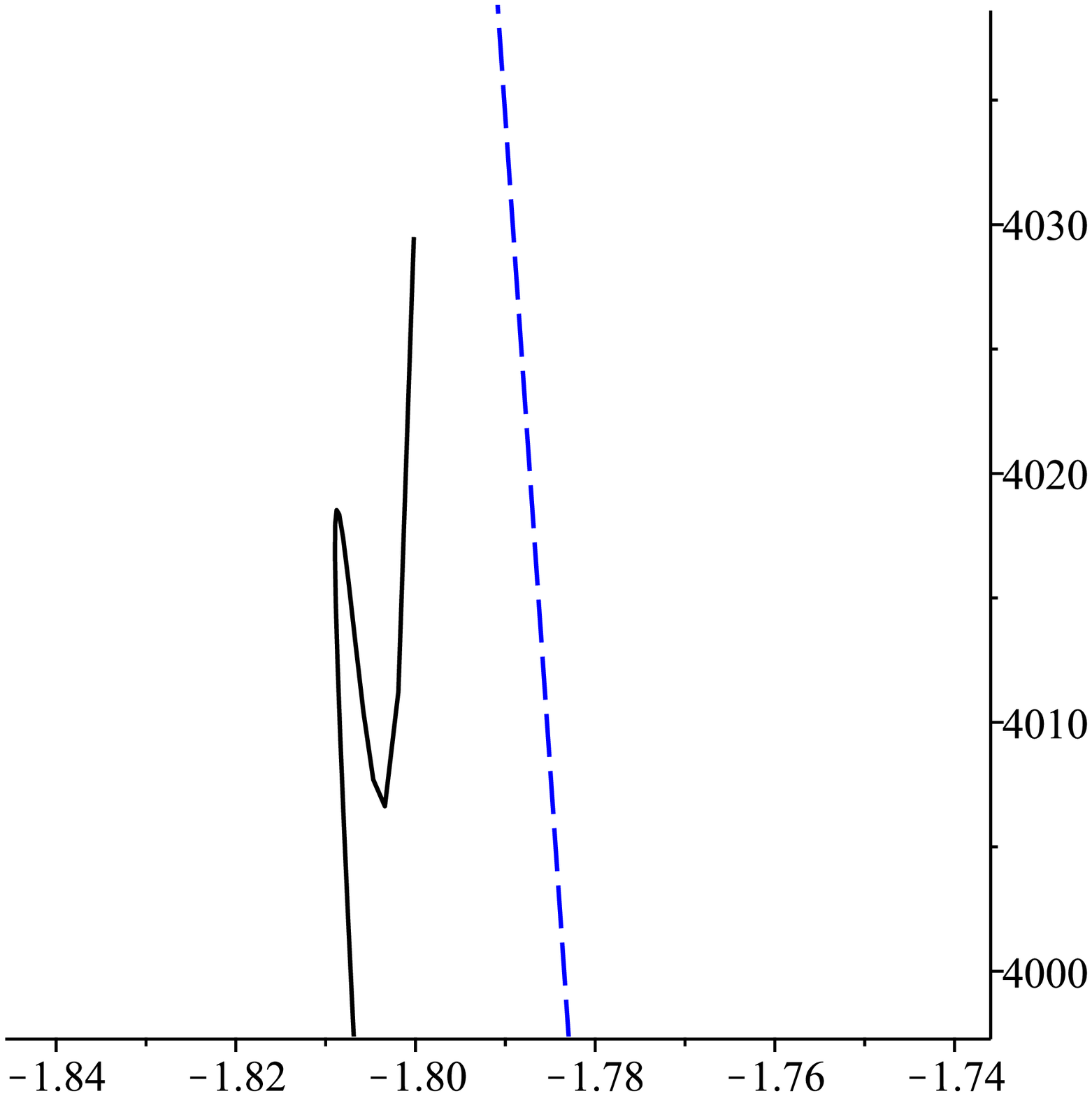}\\
        \rule{0ex}{0.85in}%
      }
  \rule{0.1in}{0ex}
  }
\small (b)
  \end{tabular}
  \begin{tabular}[b]{c}
  \includegraphics[scale=0.25]{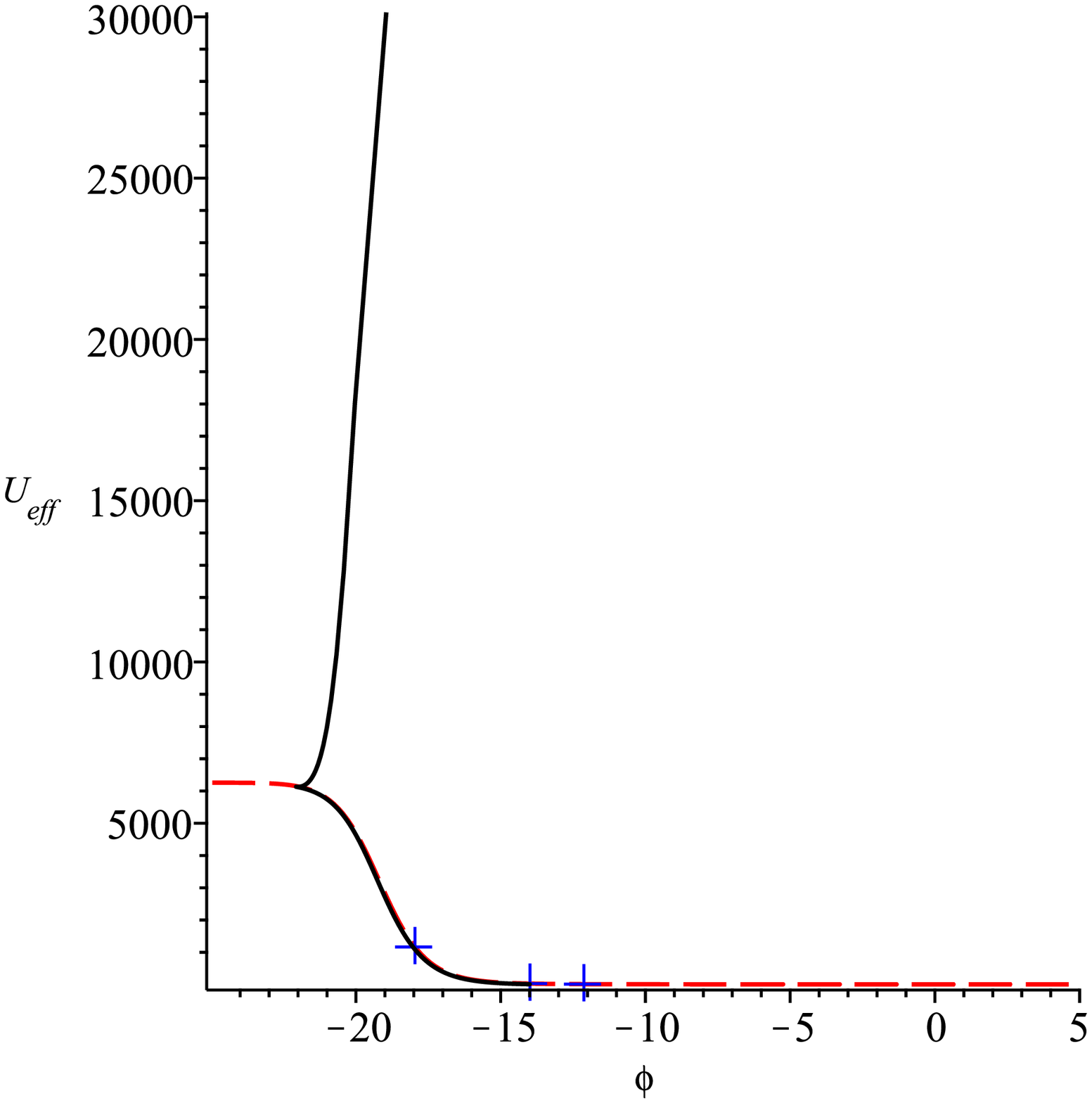}
\centering
\llap{\shortstack{%
        \includegraphics[scale=.1275]{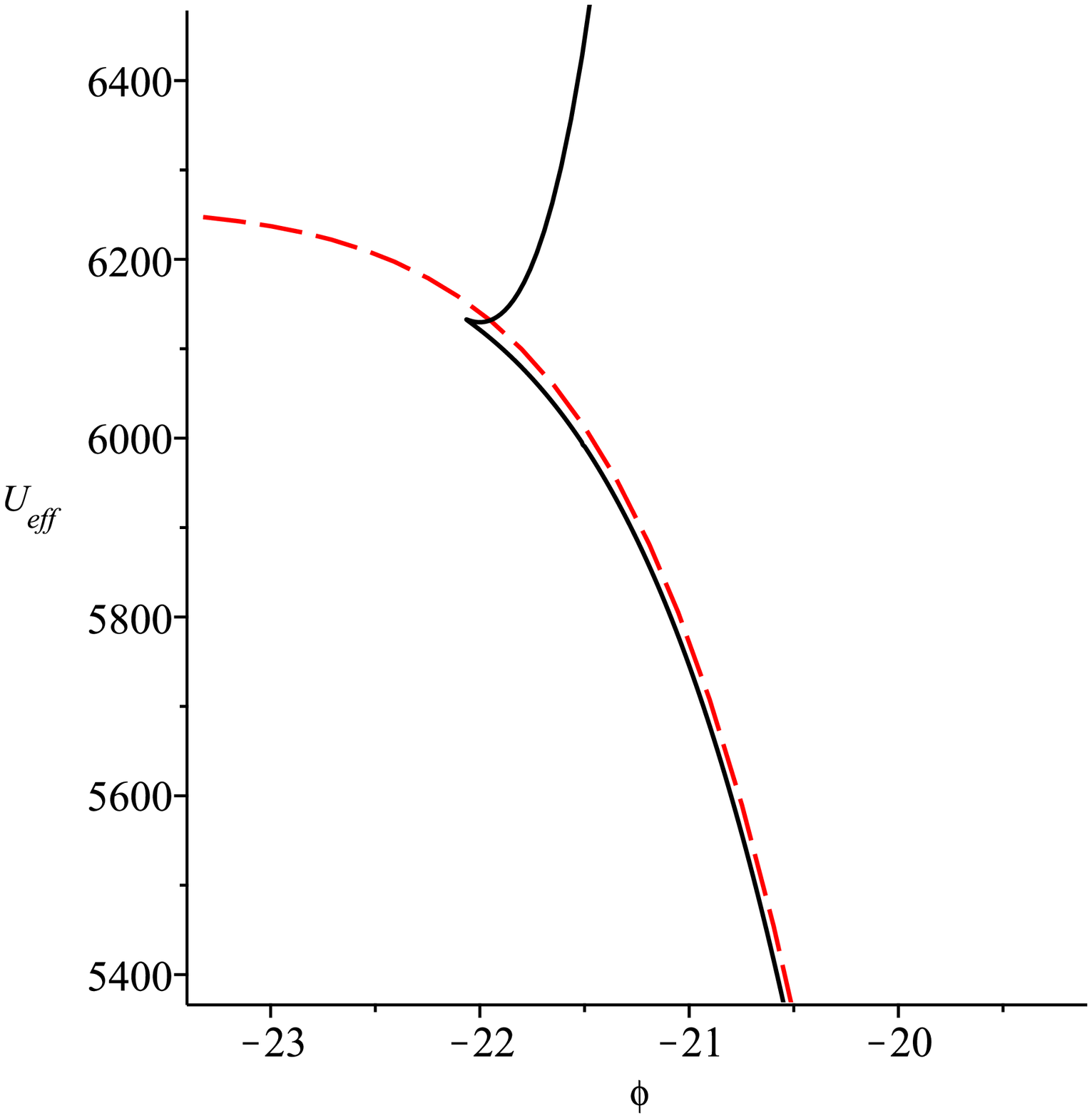}\\
        \rule{0ex}{0.8in}%
      }
  \rule{0.1in}{0ex}}
  \small (c)
  \end{tabular}
  \caption{a) Dashed line -- case 1 $(\chi_2=1,\;M_0=-0.04,\;M_1=1.5,\;M_2=0.001)$
   for parameters $\{\alpha,b_0, p_u, f_1,f_2\}$ =  $\{1, 0.027, 7.7\times10^{-9}, 7, 10^{-3}\}$ .
   Solid line -- case 2 $(\chi_2=4\!\times\!10^{-5},\;M_0=-0.01,\;M_1=0.763,\;M_2=4)$. 
  for parameters $\{\alpha,b_0, p_u, f_1,f_2\}$ = 
 $\{0.64,\, 1.41\!\times\!10^{-7},\, 6.5\!\times\!10^{-24},10^{-4}, 10^{-8}\}$ . With crosses are the points $t=0, t=$ and $t=2$, with diamonds the point $\phi_{max}$. b) and c) The  numerical integration of $T_5$ with respect to $\phi$ in Case 1 and 2, respectively denoted with solid lines. The effective potential is with dashed lines. The crosses are the same as on a) One can see that while the effect from the increase of absolute value of $\phi$ in the beginning is much more dramatic in Case 2, it also appears in Case 1.       
}  
\label{Fig5}
\end{figure}

From those observation it seems that the effective potential cannot be trusted entirely with respect to the effect of ``climbing up the slope''. To put some light on it, we attempt to reconstruct the effective potential $U_{eff}^{num}$ from $T_5$ trough the means of numerical integration. We do that based on the assumption that the effective potential from equation (\ref{U_eff}) is an approximation to $\int{T_5(\phi)}d\phi$, which seems to be confirmed by the nice coincidence between the two at most of the times. Because the function $T_5$ is not explicit function of only $\phi$, we use numerical integration of the data points $[\phi(t_i), T_5(t_i)]$. To do that we use a modified Simpson's rule \cite{Simpson} adapted to work in Maple, using 1000 datapoints, which seems sufficient to cover well the interval. The constant of integration has been determined by comparing the numerirical potential and the effective one at late times. 

The preliminary results are plotted on Figure \ref{Fig5}. On Figure \ref{Fig5} a) we illustrate the so-called ``climibing up the slope'', using the effective potentials (Equation \ref{U_eff}) for Case 1 and 2. The crosses on each one show the moments $t=0,1,2$, the diamonds mark the maximal value of $\phi(t)$. While in Case 1, this point is just slightly above the initial cross, in Case 2, it seems to reach the plateau of the potential, which is a zone that we have ruled out as nonphysical in \cite{1806.08199}.   

On Figure \ref{Fig5} b) and c) we have plotted the reconstructed numerically from $T_5$ potential.  One can see that while the two potentials coincide at later times (or their corresponding $\phi$), there is a serious deviation between the effective potential and the numerical potential during early times. While the interval in $t$ where $T_5\neq U_{eff}'$ is very small, the interval in $\phi$ is much larger because $\phi$ changes significantly in this time-interval. Furthermore, the shape of the potential term (solid line on the plots) is much more complicated than the theoretical one (dashed line), it does not posses a left plateau at all and instead it is extended in a non-trivial way. It shows that there is no ``climbing up the potential'' but instead, the potential has a complicated form with a local maximum and minimum over which the inflaton is rolling down. One can see also that in Case 1, the ``added'' part of the potential is much lower than the one in Case 2 and the local maximum (i.e. the point where the numerical potential meets the effective one) is higher. One can connect the shorter and weaker initial inflation in Case 1 with that particular shape of the potential (in Case 1, the inflaton has to go trough a potential well, starting from a much lower point, while in Case 2, it starts from much higher and the potential well is almost not existing). 

Despite this numerical result being very interesting, one needs to emphasize that describing the behaviour of the system in terms of classical quantities like the potential and the kinetic terms is difficult, because the inflaton equation (\ref{inflaton_eq}) during the earliest time, has much more complicated form than the standard equation of motion (\ref{can}), with strong dependence on both the darkon and the inflaton and their velocities. For this reason, the so presented numerical results have to be further investigated analytically.

\section{CONCLUSION}
We have presented some novel properties of the multi-measure model with two scalar fields -- the inflaton and the darkon. We have studied the areas of validity of the effective potential approximation and we have shown, that only in one of the studied cases the inflaton equation can be reduced to the standard form of equation of motion of the inflaton featuring kinetic and potential term. In the general case, it is not possible to simplify significantly the equation and thus the traditional description of the motion of the inflaton as rolling down a steep potential may be flawed. 

We have numerically studied the potential term and we have shown that after some moment in time, the effective potential indeed a very good approximation of the potential term of the inflaton equation and the only deviation is seen during the early times of the integration. 

We have also demonstrated a new property of the evolution of the system -- namely, an inflaton increasing its absolute value (or the so called ``climbing up the effective potential''). This new property allows for constant-rate inflation and allows us to overcome the problem of weak exponentiality during early inflation. To investigate it further, we have numerically reconstructed the effective potential from the actual potential term and we have seen that in fact, the small deviation during early times, leads to a huge deviation from the expected effective potential as a function of the inflaton field $\phi$. The so-obtained numerical effective potential has additional local maximum and minimum, which may explain why the two considered cases have such different properties.  The recovered shape of the potential demonstrates the advantage of using exact numerical methods to supplement analytical approximations, especially when those approximations may not be numerically justified or when it is difficult to gain physical intuition on the problem due to its large parameter-space. 

\section{ACKNOWLEDGMENTS}
It is a pleasure to thank E. Nissimov and S. Pacheva for the discussions, as well as M. Stoilov for his useful comments. 

The work is supported by BAS contract DFNP -- 49/21.04.2016, by Bulgarian NSF grant DN-18/1/10.12.2017 and by Bulgarian NSF grant 8-17.

\nocite{*}
\bibliographystyle{aipnum-cp}%

\begin{thebibliography}{}
%
%

\bibitem{ref01} E.I. Guendelman, Mod. Phys. Lett. A {\bf 14} 1043-1052 (1999), arXiv:gr-qc/9901017;
\bibitem{ref01_1} E. Guendelman  and  A.  Kaganovich, Phys.  Rev. D {\bf 60} 065004 (1999), arXiv:gr- qc/9905029; 
\bibitem{ref01_2} E. Guendelman and O. Katz,  Class. Quantum Grav. {\bf 20} 1715-1728 (2003), arXiv:gr-qc/0211095

\bibitem{ref01_3} E.I. Guendelman and P. Labrana, Int. J. Mod. Phys. D22 (2013) 1330018, arxiv:1303.7267 [astro-ph.CO];
E.I. Guendelman, D. Singleton and N. Yongram, JCAP 1211 (2012) 044, arxiv:1205.1056 [gr-qc];
E.I. Guendelman, H.Nishino and S. Rajpoot, Phys. Lett. 732B (2014) 156, arXiv:1403.4199 [hep-th].

\bibitem{1407.6281} E.I. Guendelman, E. Nissimov and S. Pacheva, {\it Unification of Inflation and Dark Energy from Spontaneous Breaking of Scale Invariance}, "Eight Mathematical Physics Meeting'', 
pp.93-103, B. Dragovic and I. Salom (eds.), Belgrade Inst. Phys. Press (2015) {\bf 10}, arXiv:1407.6281 [hep-th]
%
\bibitem{1408.5344}E. Guendelman, R. Herrera, P. Labrana, E. Nissimov, S. Pacheva, {\it Emergent Cosmology, Inflation and Dark Energy}, General Relativity and Gravitation {\bf 47} (2015) 
art.10, arXiv:1408.5344 [gr-qc]
%
\bibitem{1507.08878}E. Guendelman, R. Herrera, P. Labrana, E. Nissimov, S. Pacheva, {\it Stable Emergent Universe -- A Creation without Big-Bang}, Astronomische Nachrichten {\bf 336} (2015) 810-814, 
arXiv:1507.08878 [hep-th]
%
\bibitem{1603.06231} E. Guendelman, E. Nissimov, S. Pacheva, {\it Gravity-Assisted Emergent Higgs Mechanism in the Post-Inflationary Epoch}, International Journal of Modern Physics D {\bf 
25} 
(2016) 1644008,  arXiv:1603.06231 [hep-th]
%
\bibitem{1609.06915} Eduardo Guendelman, Emil Nissimov, Svetlana Pacheva.,{\it Quintessential Inflation, Unified Dark Energy and Dark Matter, and Higgs Mechanism},  Bulgarian Journal of Physics 44 
(2017) 15-30, arXiv:1609.06915[gr-qc] 

\bibitem{cosmo0} L. Marochnik (2016) {\it  Dark Energy and Inflation in a Gravitational Wave Dominated Universe}  	Gravitation and Cosmology, {\bf 22}, 10, 2016,   arXiv:1508.07312
[physics.gen-ph] 
\bibitem{Linde} A. Linde, {\em Inflationary Cosmology after Planck 2013}, arXiv:1402.0526 [hep-th]

\bibitem{Debono} I. Debono, G. F. Smoot, General Relativity and Cosmology: Unsolved Questions and Future Directions, Universe. 2016; 2(4):23
%
\bibitem{1610.08368} D. Staicova, M. Stoilov, {\em Cosmological Aspects Of A Unified Dark Energy And Dust Dark Matter Model}, Mod. Phys. Lett. A, Vol. 32, No. 1 (2017) 1750006, arXiv:1610.08368 
%
\bibitem{1801.07133} D. Staicova, M. Stoilov, {\em Cosmological solutions from models with unified dark energy and dark matter and with inflaton field }, arXiv: 1801.07133
\bibitem{1806.08199} D. Staicova, M. Stoilov, {\em Cosmological solutions from multi-measure model with inflaton field }, arXiv: 1806.08199
%
\bibitem{1411.5021}  H.  Motohashi,  A.  A.  Starobinsky,    and  J.  Yokoyama, {\em Inflation with a constant rate of roll}, JCAP 1509, 018 (2015), arXiv:1411.5021 [astro-ph.CO].
%
\bibitem{1702.05847} H.  Motohashi  and  A.  A.  Starobinsky, {\em Constant-roll inflation: confrontation with recent observational data},  EPL 117,  39001 (2017), arXiv:1702.05847 [astro-ph.CO]
%
\bibitem{1808.01325} Jose T. Galvez Ghersi, Alex Zucca, Andrei V. Frolov, {\em Observational Constraints on Constant Roll Inflation}, arXiv:1808.01325 [astro-ph.CO] (2018)

%
%
\bibitem{9408015} A. R. Liddle, P. Parsons, J. D. Barrow, {\em Formalising the Slow-Roll Approximation in Inflation},  	Phys.Rev. D50 (1994) 7222-7232,  	arXiv:astro-ph/9408015
\bibitem{Simpson} L. V. BLAKE, {\em A Modified Simpson's Rule and Fortran Subroutine for Cumulative Numerical Integration of a Function Defined by Data Point},  	NHL
Memorandum Report 2231 (1971),  http://www.dtic.mil/dtic/tr/fulltext/u2/723583.pdf


\end{thebibliography}

\end{document}